\documentclass[
reprint,
superscriptaddress,
showpacs,
nofootinbib,
amsmath,amssymb,
aps,
prl,
floatfix,longbibliography
]{revtex4-2}
\usepackage{graphicx}
\usepackage{dcolumn}
\usepackage{float}
\usepackage{booktabs}
\usepackage[hypertexnames, breaklinks=true, bookmarksnumbered=true, bookmarksopen=true, colorlinks=true, linktocpage=true, citecolor=blue, urlcolor=magenta, linkcolor=magenta]{hyperref}
\usepackage{amsmath,amssymb,amsfonts,amsthm}
\usepackage{mathtools}
\usepackage[T1]{fontenc}
\usepackage[latin9]{inputenc}
\usepackage{txfonts}
\usepackage{bbm}
\usepackage{bm}
\usepackage{mathrsfs}
\usepackage{xcolor}
\usepackage{natbib}
\usepackage{siunitx}

\usepackage{appendix}
\begin{document}

\title{Collective quantum stochastic resonance in Rydberg atoms}

\author{Haowei Li}
\thanks{These authors contributed equally to this work}
\affiliation{CAS Key Laboratory of Quantum Information, University of Science and Technology of China, Hefei 230026, China}
\author{Konghao Sun}
\thanks{These authors contributed equally to this work}
\affiliation{CAS Key Laboratory of Quantum Information, University of Science and Technology of China, Hefei 230026, China}
\author{Wei Yi}
\email{wyiz@ustc.edu.cn}
\affiliation{CAS Key Laboratory of Quantum Information, University of Science and Technology of China, Hefei 230026, China}
\affiliation{CAS Center For Excellence in Quantum Information and Quantum Physics, Hefei 230026, China}
\affiliation{Hefei National Laboratory, University of Science and Technology of China, Hefei 230088, China}

\begin{abstract}
We study the collective response of a group of dissipative Rydberg atoms to a periodic modulation of the Rydberg excitation laser.
Focusing on the emergent collective-jump dynamics, where the system stochastically switches between states with distinct Rydberg excitations, we show that the counting statistics of the state switching is qualitatively changed by the periodic drive.
The impact is most prominent when the driving frequency is comparable to the emergent collective-jump rate, as the jumps tend to synchronize with the external drive, and their counting statistics exhibits a series of suppressed subharmonics of the driving frequency.
These phenomena are manifestations of a novel type of stochastic resonance, where a cooperative collective state switching is facilitated by quantum fluctuations in a many-body open system.
Such a collective quantum stochastic resonance further leads
to an enhanced signal-to-noise ratio in the power spectrum of the Rydberg excitations,
for which the synchronized collective jumps are viewed as the output signal.
We confirm the many-body quantum nature of the resonance by devising
a cluster model, under which the role of many-body correlations is analyzed by changing the size of the atom clusters.
\end{abstract}

\maketitle

{\it Introduction.}
Stochastic resonance describes the counterintuitive signal amplification by adding an appropriate dose of noise~\cite{rmp70,snrdef}.
Originally suggested to explain the recurring ice ages~\cite{benzi1}, stochastic resonance proves to be a universal phenomenon in nature, appearing also in biological processes~\cite{bio1,bio2}, and across a variety of physical systems ranging from optics~\cite{opt0,opti1,opti2,amo,opti3,opti4} to electronics~\cite{elec1,elec2}.
The occurrence of stochastic resonance generally entails a noisy bistable system periodically modulated by a weak signal.
When the frequency of the driving signal and the noise strength are appropriate, the response of the system exhibits resonance-like behaviors, including the synchronization of the bistable-state switching with the external drive, and an enhanced signal-to-noise ratio (SNR).
While stochastic resonances are predominantly found in non-linear systems with classical noise, they also emerge in quantum mechanical settings, assisted by quantum fluctuations~\cite{qsr1,qsr2,qsr3,qsr4,qsr5}.
These quantum stochastic resonances have only recently been observed in quantum-dot microstructures~\cite{qdot} and scanning tunneling microscopes~\cite{microscope}, where the intrinsic fluctuations of single-electron tunneling play a key role.

But quantum fluctuations, particularly those in many-body open systems, also manifest in collective dynamics.
Take a group of dissipative Rydberg atoms for instance.
The atomic ground states therein are laser-coupled to high-lying energy levels with significant principle quantum number $n$, but subject to spontaneous decay back to the ground state.
An outstanding feature of the Rydberg atoms is the strong long-range interactions, which are central to the recent advances in quantum computation and simulation with Rydberg atoms~\cite{rydrev1, rydrev2, ryd1,ryd2,ryd3,ryd4,ryd5,ryd6,ryd7}.
Previous studies have shown that the entanglement generated by the long-range interactions, combined with the repetitive quantum measurements (in the form of spontaneous emission), lead to collective quantum jumps between two metastable many-body states with high and low Rydberg excitations, respectively~\cite{cj}.
Since these emergent collective jumps are a manifestation of quantum fluctuations in a many-body open system, it is tempting to ask whether a quantum stochastic resonance can be engineered based on them.

 In this work, we reveal such a quantum stochastic resonance in the counting statistics of the collective jumps in a group of Rydberg atoms, where the Rydberg excitation laser is subject to a periodic modulation.
While rich resonance behaviors emerge in the stochastic state evolution of the many-body open system,
many-body correlations are indispensable for the observed resonance, in contrast to previously studied stochastic resonances.
We explicitly demonstrate this using a cluster model, where quantum coherence and entanglement are only retained within each atom cluster. Specifically, the cluster size and partitions dramatically impact the collective jumps and resonance condition.

%


\begin{figure}[tbp]
\includegraphics[width=0.48\textwidth]{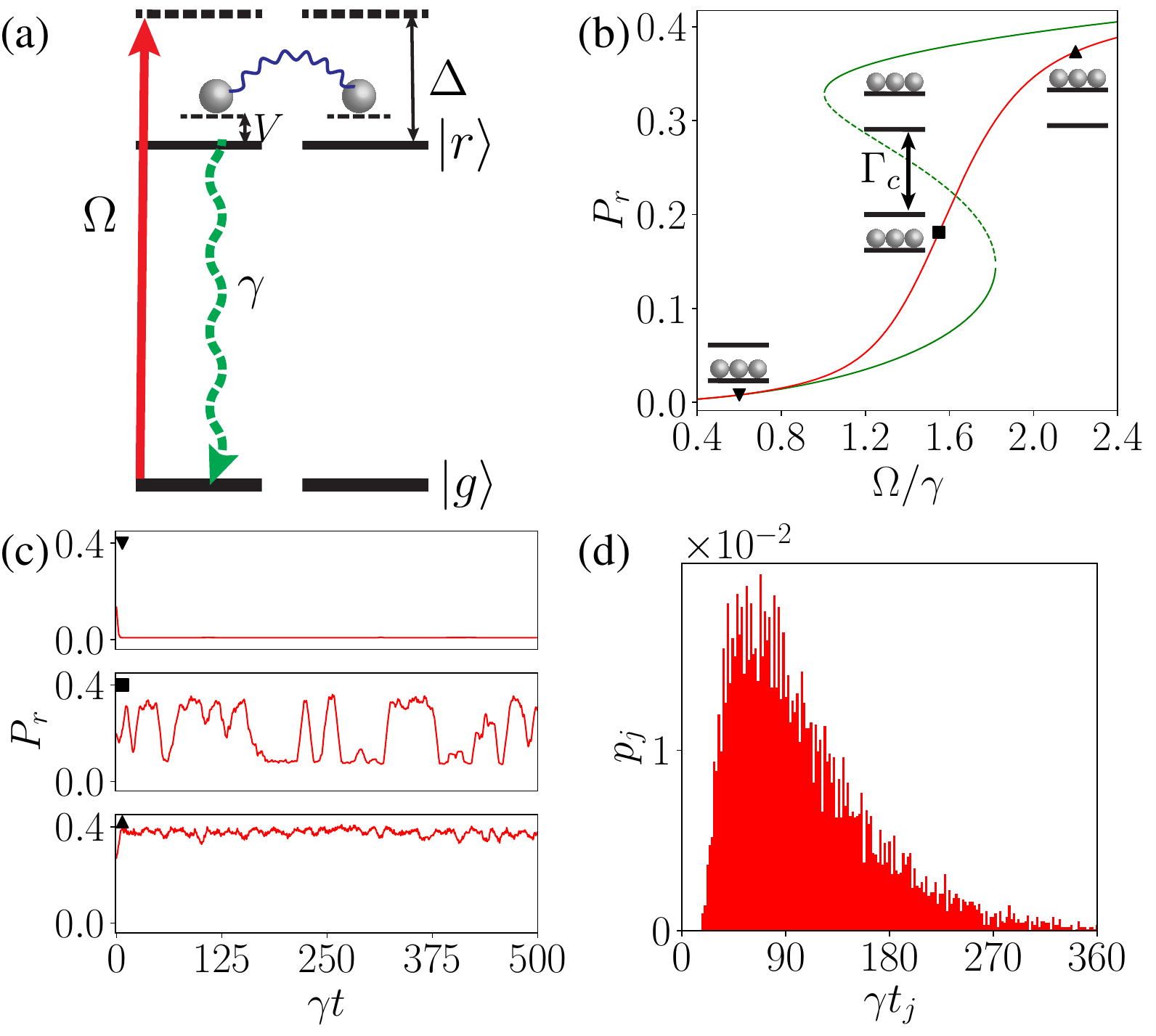}
	\caption{(a) Schematic illustration of the level scheme. The Rydberg excitation laser couples the atomic ground state $|g\rangle$ and a Rydberg state $|r\rangle$. Atoms in the excited state
interact with each other through the Rydberg long-range interaction, and can spontaneously decay back to the ground state. (b) Rydberg-state population $P_r$ in the steady state of the master equation (red), and in the bistable states of the optical Bloch equations (green).  Note that solid (dashed) sectors of the green curve correspond to stable (unstable) solutions.
Collective jumps with rate $\Gamma_c$ (defined in the main text) predominantly occur in the mean-field bistable region, but only visible through quantum trajectories. Here the mean-field bistable region exists in the range $\Omega/\gamma \in [1.00,1.82]$.
(c) Rydberg excitation along a single trajectory. From top to bottom: $\Omega/\gamma=0.6, 1.55, 2.2$, consistent with the corresponding black {markers} in (b).
(d) Normalized distribution $p_j$ of inter-jump intervals, indicative of the counting statistics of the collective jumps in the middle panel of (c). We take $\gamma\delta t=2$ for the definition of $p_j$ and $t_j$ (see main text for details).
For numerical calculations, we take $\Delta/\gamma=3.4$, $(N-1)V/\gamma=10$, and $N=8$ in all figures.}
		\label{fig1}
\end{figure}

{\it Emergent collective jumps in Rydberg atoms.}
We consider a group of $N$ atoms uniformly subject to a Rydberg excitation laser.
As illustrated in Fig.~\ref{fig1}(a), the ground state of the $j$th atom $|g_j\rangle$ is continuously coupled to a Rydberg state $|r_j\rangle$. The Rydberg state has a finite lifetime, which can be further tuned through additional couplings to other excited states. Approximating the Rydberg atoms as an ensemble of two-level systems and neglecting the much slower external motion, the dissipative dynamics of the density matrix $\rho$ is depicted by the Lindblad master equation
\begin{align}
\dot{\rho}=-i[H, \rho]+\gamma \sum_j\left(-\frac{1}{2}\{L_j^\dagger L_j, \rho\}+L_j\rho L_j^\dagger\right),
		\label{eq:lindblad}
\end{align}
with the Hamiltonian
\begin{align}
H=\sum_j\left[-\frac{\Delta}{2}(\sigma_j^z+1)+\frac{\Omega}{2}\sigma_j^x\right] +\frac{V}{4} \sum_{j<k}(\sigma_j^z+1)  (\sigma_k^z+1).\label{eq:H}
\end{align}
The quantum jump operator
$L_j=(\sigma_j^x-\mathrm{i}\sigma_j^y)/2$ describes the spontaneous decay of the $j$th atom back to its ground state, with a decay rate $\gamma$.
Here $\Omega$ is the Rabi frequency of the Rydberg excitation laser, $\Delta$ is its detuning, and $\sigma^{x,y,z}_j$ are the Pauli operators associated with the two-dimensional subspace spanned by $\{|g_j\rangle, |r_j\rangle\}$ of the $j$th atom, with $\sigma^z_j=|r_j\rangle\langle r_j|-|g_j\rangle\langle g_j|$.
For simplicity, we model the long-range Rydberg interactions using {a constant all-to-all coupling $V$}. Note that all the key results still hold when we consider the spatial dependence of the Rydberg interactions~\cite{sm}.

The driven-dissipative Rydberg system features a unique steady state (with density matrix $\rho^s$), which can be solved by setting $\dot{\rho^{s}}=0$ in Eq.~(\ref{eq:lindblad}). In Fig.~\ref{fig1}(b), we show the steady-state Rydberg excitation $P_r=\rho^{\text{s}}_{rr}$ for $N=8$ atoms (red). For later discussions, it is instructive to compare the steady-state solution from the master equation with those under a mean-field approximation. In the latter case, the interatomic correlations are neglected, the density matrices of individual atoms become decoupled through $\rho=\bigotimes_j \bar{\rho}$, and the dynamics of $\bar{\rho}$ is
governed by the optical Bloch equations~\cite{mf,cj}
	\begin{align}
\dot{\bar{\rho}}_{r r}&=-\Omega \operatorname{Im} \bar{\rho}_{r g}-\gamma \bar{\rho}_{r r}, \label{eq:op1}\\
\dot{\bar{\rho}}_{r g}&=i\left[\Delta-(N-1)V \bar{\rho}_{r r}\right] \bar{\rho}_{r g}-\frac{\gamma}{2} \bar{\rho}_{r g}+i \Omega\left(\bar{\rho}_{r r}-\frac{1}{2}\right),\label{eq:op2}
	\end{align}
where the Rydberg interactions contribute toward a density-dependent effective detuning.
Steady-state solutions can be obtained by setting $\dot{\bar{\rho}}=0$ in the equations above,
giving rise to the green lines in Fig.~\ref{fig1}(b).
Notably, a bistable region emerges, where the two steady-state solutions feature high and low Rydberg excitations, respectively. The bistability derives from the non-linear interaction terms
under the mean-field approximation, and is well-studied in thermal Rydberg gases~\cite{therm1,therm2,therm3}. While such a bistability is absent in the full-quantum description of Eq.~(\ref{eq:lindblad}),
its quantum correspondence nevertheless manifests in the counting statistics of the open-system dynamics.

\begin{figure}[tbp]
	\includegraphics[width=0.48\textwidth]{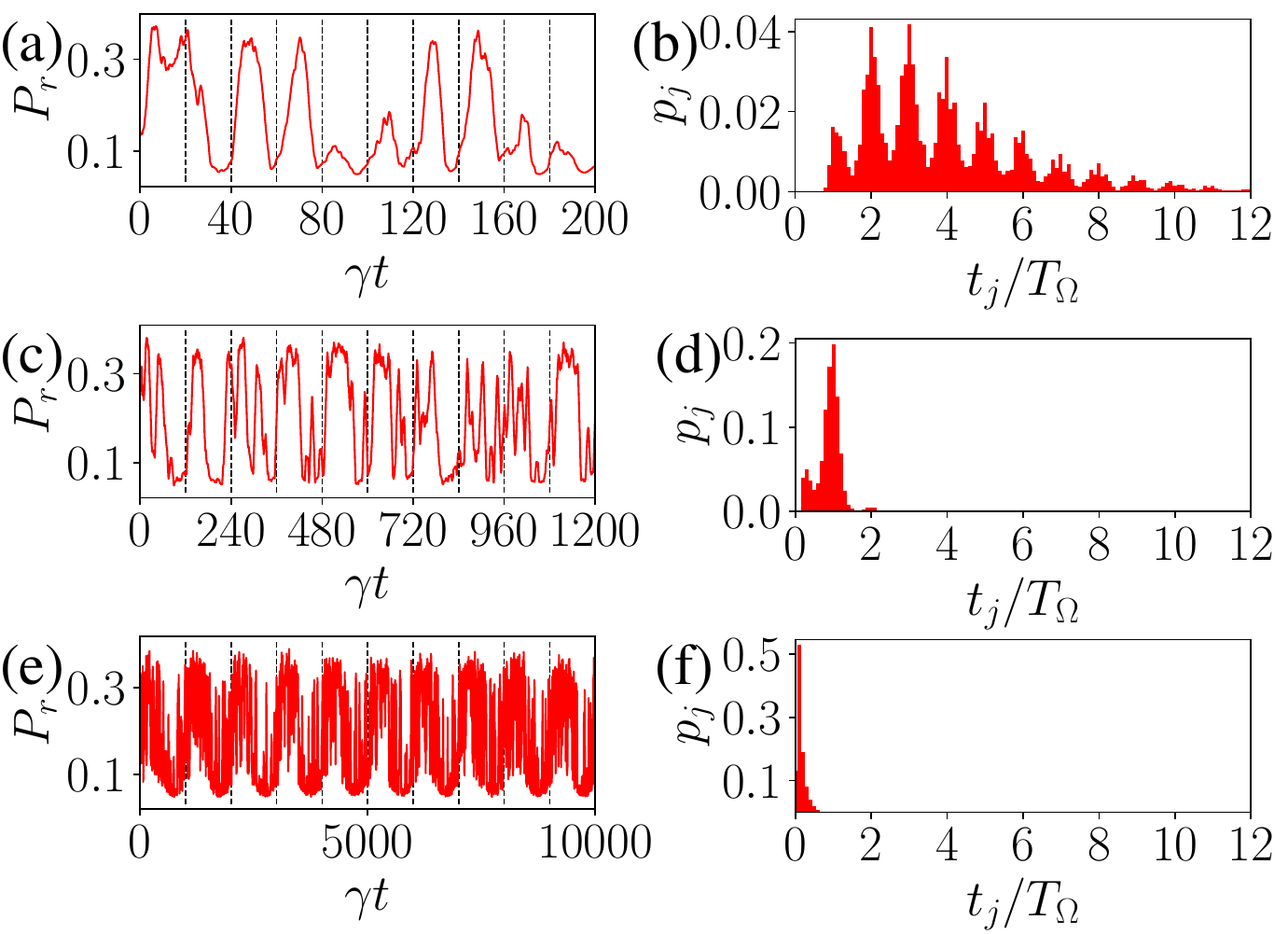}
	\caption{(a)(c)(e) Single-trajectory dynamics of the Rydberg excitation $P_r$ under a periodic drive.
		(b)(d)(f) Distribution of inter-jump intervals for a single trajectory within the evolution time $\gamma t\in [0, 10^6]$. The driving parameters are $\Omega_0/\gamma=1.55$, $A_\Omega/\gamma=0.2$, and (a)(b) $\gamma T_\Omega=20$, (c)(d) $\gamma T_\Omega=120$, (e)(f) $\gamma T_\Omega=1000$. For all figures, we take $N=8$ and $\delta t=0.1T_\Omega$. Other parameters are the same as those in Fig.~\ref{fig1}.}
	\label{fig2}
\end{figure}

To see this, we unravel the density-matrix dynamics into an ensemble of quantum trajectories~\cite{traj1,traj2,traj3}. In a given time interval $[t,t+\Delta t]$ along each trajectory, the many-body wave function $|\psi(t)\rangle$ undergoes the quantum jump $|\psi(t)\rangle\rightarrow L_j|\psi(t)\rangle$ with a probability $\gamma\Delta t \langle \psi(t)|L_j^\dag L_j|\psi(t)\rangle$, or evolves under a non-Hermitian effective Hamiltonian $H_{\text{eff}}=H-i\frac{\gamma}{2}\sum_{j}L^\dag_j L_j$ with the probability $1-\gamma\Delta t \sum_j \langle \psi(t)|L_j^\dag L_j|\psi(t)\rangle$. The state is then normalized to repeat the process for the following time interval.
 Given our unraveling scheme above, each single trajectory simulates the many-body wave-function evolution in a single-run experiment, provided the spontaneously emitted photons are directly detected~\cite{milburn92,gp92p,gp92j,sm}.

An important feature of an individual quantum trajectory of the system is the possible emergence of collective jumps between metastable many-body states with high and low Rydberg excitations~\cite{cj}. The two states are characterized by the joint emissions or non-emissions of individual atoms, and the collective jumps between them originate from the interplay of Rydberg interaction and dissipation.
In Fig.~\ref{fig1}(c), we show the single-trajectory evolution of the Rydberg excitation $P_r=\frac{1}{2}\langle\psi(t)|\sum_j(\sigma^z_j+1)|\psi(t)\rangle$ with different $\Omega$.
Apparently, when the parameters are within the mean-field bistable region [Fig.~\ref{fig1}(c) middle panel], $P_r$ switches between two plateaus, whose values correspond to the bistable steady-state solutions under the mean-field approximation.
When the system is tuned away from the bistable region, the collective jumps dramatically decrease in frequency, and eventually vanish in effect [Fig.~\ref{fig1}(c) top and bottom panels]. This is the case when the steady-state solution of the master equation is close to one of the steady-state solutions of the optical Bloch equations.

To characterize the counting statistics of the collective jumps, we plot the distribution of inter-jump intervals in Fig.~\ref{fig1}(d).
We uniformly discretize the overall evolution time $T$ into small segments of $\delta t$, and count the number of adjacent downward collective jumps $n_j$ that occur within the time interval $[t_j-0.5\delta t,t_j+0.5\delta t]$, where $t_j=j\delta t$.
Here the downward collective jumps are the state switching from high to low Rydberg excitations.
 The time of occurrence of these downward jumps along a quantum trajectory is determined numerically by setting a population-difference threshold $P^{\text{th}}_r=0.25$~\cite{sm}.
The normalized distribution for the inter-jump interval is defined as $p_j=n_j/\sum_j n_j$,
which represents the probability for adjacent downward collective jumps to occur within the corresponding time interval. The distribution shows an exponential decay for large $j$, suggesting the stochastic nature of the jumps.
We further define a collective-jump rate $\Gamma_c=\sum_j n_j/T$, which represents an emergent many-body energy scale for the driven-dissipative system.

{\it Counting statistics and quantum stochastic resonance.}
We now consider the counting statistics of the collective jumps in response to a periodic modulation of the Rydberg excitation laser,
with $\Omega(t)=\Omega_0+A_\Omega \cos(2\pi t/T_{\Omega})$, where $T_\Omega$ and $A_{\Omega}$ are respectively the driving period and amplitude, and $\Omega_0$ is the time-independent Rabi frequency.

In Fig.~\ref{fig2}(a)(c)(e), we show the time-evolved Rydberg excitations of single trajectories under different driving frequencies.
The collective jumps appear to
be regulated by the driving frequency, distinct from the more random pattern in Fig.~\ref{fig1}(c).
Importantly, at an appropriate intermediate driving frequency, the collective jumps fully synchronize with the drive [see Fig.~\ref{fig2}(c)].
Such a pattern is more visible from the counting statistics of the jumps, shown in Fig.~\ref{fig2}(b)(d)(f).
When the driving period is small, the response of the system, in terms of the downward collective-jump intervals, features a series of subharmonics of the driving frequency [see Fig.~\ref{fig2}(b)]. On increasing the driving period, the subharmoincs become suppressed, and the distribution is peaked at $t_j/T_\Omega=1$, indicating full synchronization [see Fig.~\ref{fig2}(d)].
Under a very long driving period, more than one jumps can occur within one driving period, and $p_j$ is peaked at $t_j/T_\Omega=0$ [see Fig.~\ref{fig2}(f)]. In all cases, the distribution of $p_j$ exhibits discrete peaks centered at integer $t_j/T_\Omega$, in sharp contrast to the pattern in Fig.~\ref{fig1}(d).

The synchronization of the system's response with the periodic drive is reminiscent of the stochastic resonance, where the resonance condition is qualitatively understood as the matching between the external driving frequency and the rate of the emergent collective jumps~\cite{rmp70,qdot,rescon}. To confirm such an observation, we calculate the SNR of the collective jumps, which is an outstanding quantifier for the stochastic resonance.
We view the period modulation as a weak input signal, and the collective jumps the output. The SNR can be extracted from the power spectrum $P(f)$ of the collective jumps in a single trajectory, where $P(f)$ is the square of the Fourier transform of $P_r(t)$. The SNR is defined as~\cite{snrdef}
\begin{align}
\text{SNR}_{\text{dB}}=10\log_{10}\left[\frac{P(f=\frac{1}{T_{\Omega}})}{\frac{1}{f_{\text{max}}-f_{\text{min}}} \int_{f_{\text{min}}}^{f_{\text{max}}}P(f) df}\right],
\end{align}
where the frequency range $[f_{\text{min}},f_{\text{max}}]$ of the noise-power integral is typically taken close to but not over the driving frequency.
To compensate for the arbitrariness in the choice of $f_{\text{min},\text{max}}$, at any given point, we take three different sets of ranges: $[f_{\text{min}},f_{\text{max}}]\in (1/T_{\Omega}) \{[0.25,0.75],\,\,[1.25,1.75], \,\, [2.25, 2.75]\}$, and retain the one with the largest integrated noise power.

\begin{figure}[tbp]
\includegraphics[width=0.48\textwidth]{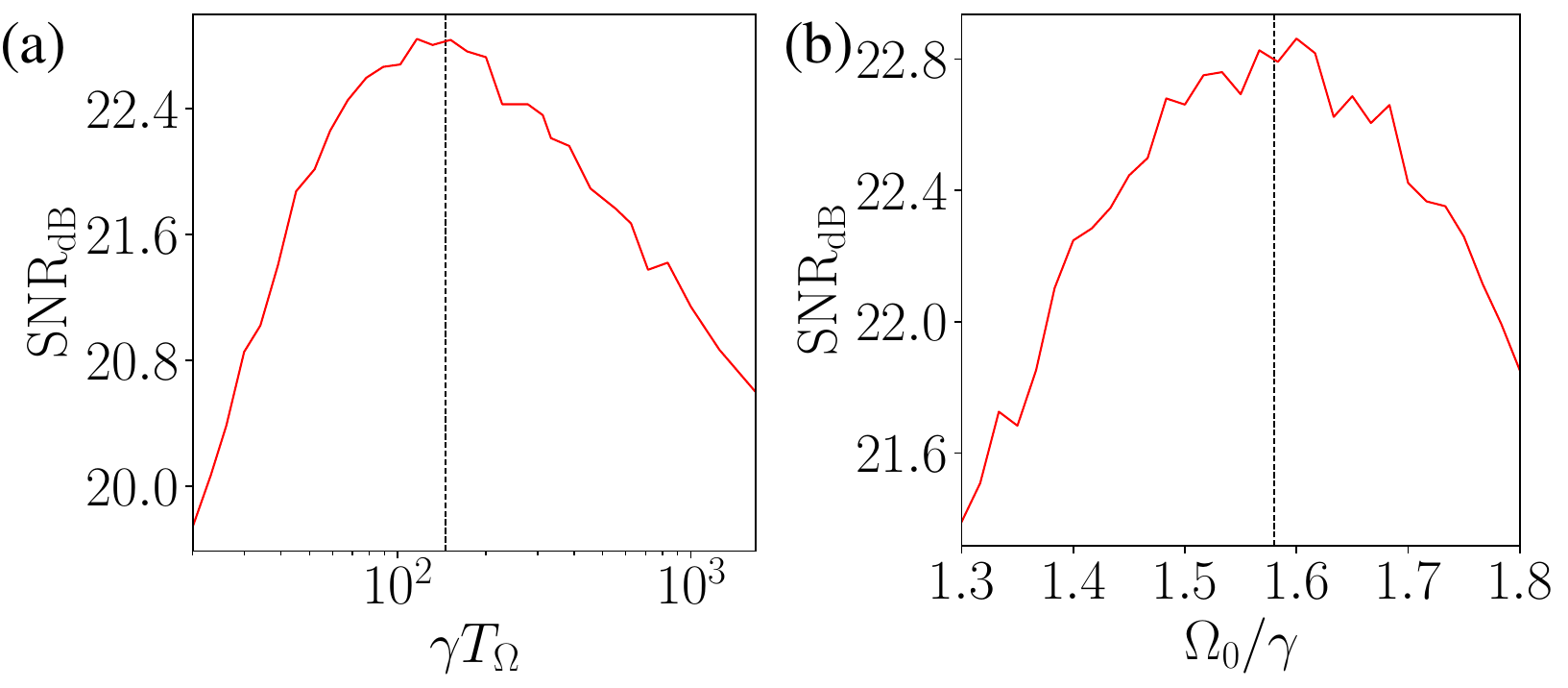}
\caption{(a) Trajectory-averaged SNR, with a fixed $\Omega_0/\gamma=1.55$, $A_\Omega/\gamma=0.2$, and varying $T_\Omega$. (b) SNR as a function of $\Omega_0$ with fixed $T_\Omega=144.2$ and $A_\Omega/\gamma=0.2$. The vertical dashed lines in both subplots indicate the
theoretically predicted resonance condition $T_c\tilde{\Gamma}_c=1$, which gives $
\gamma T_c\approx 145.4$ in (a), and $\Omega_0/\gamma\approx 1.58$ in (b).
Here $\tilde{\Gamma}_c$ is the downward collective-jump rate that is numerically averaged along the driving path~\cite{sm}.
For both calculations here, we take a single-trajectory evolution time of $\gamma t=10000$, and
average over $100$ trajectories.
Other parameters are the same as those in Fig.~\ref{fig1}.
}
		\label{fig3}
\end{figure}

We plot the numerically evaluated SNR in Fig.~\ref{fig3}(a), as a function of the driving period,
where the SNR is further averaged over many trajectories for better convergence within the numerically feasible evolution time.
The resonance point is identified as the peak location in the SNR, which
is close to the theoretically predicted resonance point $T_c \tilde{\Gamma}_c=1$~\cite{rmp70,rescon} (vertical dashed line). Here $\tilde{\Gamma_c}$ is estimated using the averaged collective-jump rate along the driving path $\Omega(t)$~\cite{sm}.
The general behavior of the SNR can be understood in view of the distribution of Fig.~\ref{fig2}(b)(d)(f).
Specifically, as the driving period increases, the subharmonics in the distribution become suppressed, and the output signal (the peak near $t_j/T_\Omega=1$) increases in weight in the response.
The SNR peak is reached when the collective jumps are fully synchronized with the drive, as a dominant peak emerges near $t_j /T_\Omega=1$.
When the driving frequency is tuned past the resonance, multiple collective jumps can occur within a single driving period, leading to a deterioration of the signal and the SNR.
Similar SNR behavior can be observed by fixing the driving period and tuning $\Omega_0$ [see Fig.~\ref{fig3}(b)]. This is equivalent to changing the intrinsic collective-jump rate to meet the resonance condition (indicated by the vertical dashed line).
As such, emergent collective jumps assist the periodic drive to enhance the SNR in an open system, which is but the hallmark of the stochastic resonance.

{\it Impact of many-body correlations.}
To understand the many-body nature of the collective quantum stochastic resonance, we consider a cluster model where interactions are treated exactly within the finite-size clusters, while mean-field approximations are adopted for inter-cluster interactions.
For simplicity, we assume all clusters to have the same number of atoms $M$ (dubbed the cluster size).

\begin{figure}[tbp]
\includegraphics[width=0.48\textwidth]{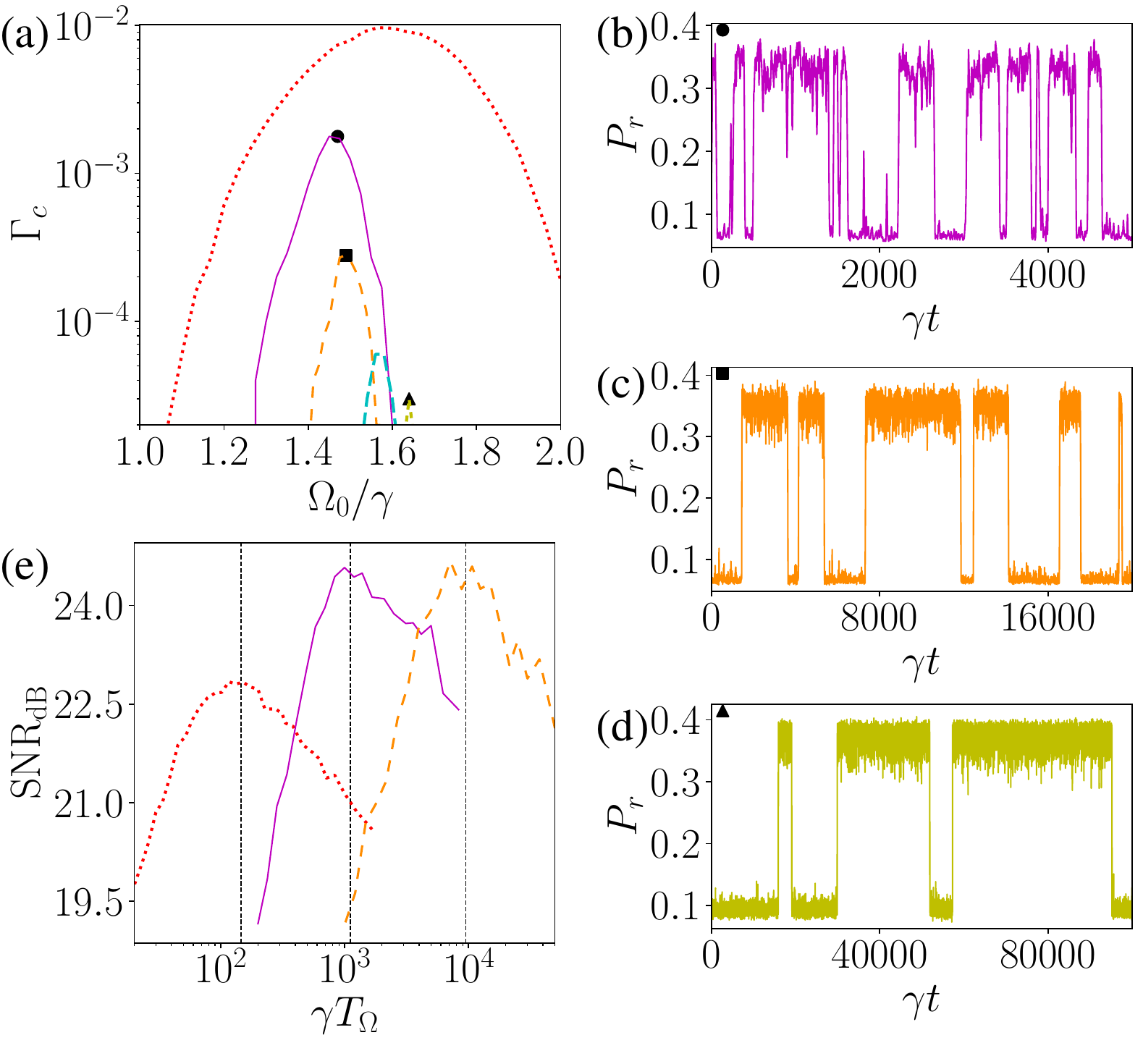}
	\caption{(a) Intrinsic downward collective-jump rates as functions of $\Omega_0$ under the cluster model in the absence of the periodic drive. The calculations are for a system of $N=8$ atoms, with $M=8$ (red), $M=4$ (purple), $M=2$ (blue), and $M=1$ (yellow), respectively.  The orange dashed curve corresponds to
a partition of three clusters, with $3$, $3$, and $2$ atoms, respectively [dubbed the $(3,3,2)$ configuration].
(b)(c)(d) Single-trajectory dynamics for (b) $M=4$ at $\Omega_0/\gamma=1.47$, (c)  the $(3,3,2)$ configuration at $\Omega_0/\gamma=1.49$, and (d) $M=1$ at $\Omega_0/\gamma=1.64$. The parameter $\Omega_0$ is chosen at the position with the maximum $\Gamma_c$ under any given $M$.
(e)  SNRs under the cluster model of $M=8$ (red dotted), $M=4$ (purple solid) and the $(3,3,2)$ configuration (orange dashed).
For $M=4$ [the (3,3,2) configuration], we fix $\Omega_0/\gamma=1.47$ ($\Omega_0/\gamma=1.49$), $A_\Omega/\gamma=0.1$ ($A_\Omega/\gamma=0.08$), and average the SNR over 35 (12) trajectories, where the evolution time for a single trajectory is $\gamma t=50000$ ($\gamma t=300000$). The vertical dashed lines indicates the theoretical resonance points, with $\gamma T_c\approx 1118.4$ and $\gamma T_c\approx 9525.2$ for the two cases, respectively.
Other parameters are the same as those in Fig.~\ref{fig1}.}
		\label{fig4}
\end{figure}

The overall density matrix can then be formally written as a direct product $\rho=\bigotimes_n \rho_n$, where $\rho_n$ is the density matrix of the $n$th cluster.
Apparently, quantum coherence and entanglement only exist within the same cluster.
While the master equation is formally the same as Eq.~(\ref{eq:lindblad}),
the Hamiltonian
of the $n$th cluster is
\begin{align}
H_n&=\sum_{j=1}^{M}\left[-\frac{\Delta}{2}(\sigma_{n,j}^z+1)+\frac{\Omega}{2}\sigma_{n,j}^x\right] \nonumber\\
&+\frac{V}{4}\sum_{j<k}(\sigma_{n,j}^z+1)  (\sigma_{n,k}^z+1)+\frac{V}{4}(\sum_{m\neq n} P^m_{r})\sum_{j=1}^{M}(\sigma_{n,j}^z+1),
\end{align}
where $\sigma_{n,j}^{x,y,z}$ are the Pauli operators for the $j$th atom in the $n$th cluster, and the Rydberg excitation of the $m$th cluster is define as
$P^m_r=\frac{1}{2}\langle \sum_{j=1}^M(\sigma^z_{m,j}+1) \rangle$.
Such a construction allows us to examine the impact of many-body correlations by decreasing the cluster size $M$, as the system is sequentially broken down into local clusters.

While the cluster model recovers the full-quantum description of Eq.~(\ref{eq:lindblad}) for $M=N$, the collective jumps are dramatically modified when $M<N$.
Particularly, in the absence of the periodic modulation, the intrinsic collective-jump rate $\Gamma_c$ significantly decreases for smaller $M$, as illustrated in Fig.~\ref{fig4}(a). We also show the single-trajectory $P_r$ for different cluster sizes in Fig.~\ref{fig4}(b)(c)(d), where the slowdown of the collective-jump rate with decreasing $M$ is directly visible.
This suggests that the resonance point should also shift toward lower frequencies at smaller $M$.  We confirm this using cluster models of smaller cluster partitions in Fig.~\ref{fig4}(e).
While the SNR peaks are consistent with theoretical predictions , they
appear at larger $T_\Omega$ than the single-cluster case with $M=8$.

Finally, we note that, in principle, the stochastic resonance persists even in the $M=1$ case, albeit at a very small resonance frequency due to the scarcity of collective jumps.
However, the quantum stochastic resonance does not exist under the optical Bloch equations of Eqs.~(\ref{eq:op1}) and (\ref{eq:op2}), due to the complete lack of individual quantum jumps therein.
Specifically, a key difference between the $M=1$ cluster model and the full mean-field description is that, in the latter case, all the atoms are in the same state and described by the same density matrix. This renders sequential atomic emissions impossible, which is crucial for the buildup of a collective jump~\cite{cj}.

{\it Discussion.}
 The collective quantum stochastic resonance is readily observable in arrays of Rydberg atoms confined to optical tweezers, or in Rydberg gases at low temperatures.
 Take the Rydberg state $|70S_{1/2}\rangle$ of $^{87}$Rb in a tweezer as an example.
Under typical experimental parameters~\cite{ryd2,ryd3}, the parameters of Fig.~3(a) correspond to $\gamma=2\pi \times 1$ MHz,
 $\Delta\sim 2\pi\times 3.4$ MHz.
{ The $\gamma$ here is larger than the spontaneous decay rate, and is achieved by lasers to couple the Rydberg state to intermediate low-lying states, thereby enhancing the overall decay rate.} The resonance then occurs at $T_{\Omega}\sim 22.95$ $\mu$s for $N=8$ atoms
with an interatomic distance $\sim 9.19$ $\mu$m (where $V\sim 2\pi \times 1.43$ MHz).
As the temperature rises, the collective-jump rate decreases along with the size of local coherent clusters,
and it would become more difficult to observe the resonance behavior.
For thermal Rydberg gases whose dynamics are well-captured by the non-linear optical Bloch equations~\cite{therm1,therm2,therm3},
neither the collective jump nor the quantum stochastic resonance persists.
Nevertheless, based on the intrinsic non-linearity-induced bistability, stochastic resonances can still emerge when some form of classical noise is introduced~\cite{sm}.
The subsequent resonance behaviors would be similar in nature to those of conventional stochastic resonances, or related phenomena in non-linear non-Hermitian systems~\cite{qiusciadv}, but are fundamentally different from what we study here.
 Finally, while we focus on a finite-size system, further analysis suggests that
the resonance behavior becomes easier to detect with larger $N$~\cite{sm}.


\begin{acknowledgments}
We thank Zongkai Liu for helpful discussions. This work is supported by the National Natural Science Foundation of China (Grant No. 12374479) and the Innovation
Program for Quantum Science and Technology (Grant No. 2021ZD0301200).
\end{acknowledgments}

\end{document}


\clearpage
\begin{widetext}
\appendix

\renewcommand{\thesection}{\Alph{section}}
\renewcommand{\thefigure}{S\arabic{figure}}
\renewcommand{\thetable}{S\Roman{table}}
\setcounter{figure}{0}
\renewcommand{\theequation}{S\arabic{equation}}
\setcounter{equation}{0}

\section{Supplemental Material for ``Collective quantum stochastic resonance in Rydberg atoms''}

In this Supplemental Material, we provide details on the identification of collective jumps,
calculation of the resonance condition, discussions on the finite-range Rydberg interactions, bistability under the cluster model, the condition for stochastic resonances within a mean-field framework, as well as the impact of different unraveling schemes and finite-size effect.

\section{Identifying collective jumps}

From a single quantum trajectory, we reply on the following numerical process to identify collective jumps
and extract time interval between jumps.
Given a quantum trajectory, we denote the time of the $n$ th downward collective jump as $t^{\mathrm{cj}}_n$. We formally set $t^{\mathrm{cj}}_0=0$, and numerically determine $t^{\mathrm{cj}}_1$ as the earliest time that the first downward jump occurs. Practically, we consider a downward jump occurs at $t^{\mathrm{cj}}_1$ if there exists an intermediate time $t^m\in (t^{\mathrm{cj}}_0, t^{\mathrm{cj}}_1)$, such that
$P_r(t^m)-P_r(t^{\mathrm{cj}}_1)\geq P_r^{\mathrm{th}}$. Here $P_r$ is the Rydberg-state population, and we fix the threshold population difference $P_r^{\text{th}}=0.25$ through numerical trial and error. Having determined $t^{\mathrm{cj}}_1$, we then repeat the process above to get a series of $t^{\mathrm{cj}}_{n}$. The time interval between two adjacent downward collective jumps is then $t^{\mathrm{cj}}_{n}-t^{\mathrm{cj}}_{n-1}$, from which we get the counting statistics.
Note that different choice of the threshold $P_r^{\text{th}}=0.25$ could deteriorate the counting statistics, but does not affect the signal-to-noise ratio (SNR) calculations, nor the occurrence of the resonance.

\section{The resonance condition}

In the literature, the resonance condition for the stochastic resonance is often given as the matching between the driving frequency with the intrinsic noise-induced switching rate. In terms of the parameters defined in the main text, this condition corresponds to $T_\Omega\Gamma_c=1$. Note that we define $\Gamma_c$ as the rate of the downward collective jump, which differs from the total collective-jump rate by a factor of two. Furthermore, as demonstrated in Fig.~4(a) of the main text, $\Gamma_c$ is a function of $\Omega_0$, and changes non-monotonically as the mean-field bistable region is traversed. While this reflects the many-body nature of the collective jumps, it also means that the intrinsic collective-jump rate varies as the Rabi frequency of the Rydberg excitation laser is driven around $\Omega_0$. To account for this variation, we use a path-averaged jump rate $\tilde{\Gamma}_c$ to estimate the resonance condition. Explicitly, we define
\begin{align}
\tilde{\Gamma}_c=\frac{1}{T_\Omega}\int_0^{T_\Omega} \Gamma_c[\Omega(t)]dt.
\end{align}
We numerically evaluate $\tilde{\Gamma}_c$, and estimate the resonance positions in Fig.~3 and Fig.~4(e), which are close to the corresponding SNR peaks.

\section{Spatial dependence of the Rydberg interactions}

Throughout the main text, we model the Rydberg interactions through an all-to-all constant coupling. In reality, depending on the experimental condition. the Rydberg interactions can be dipolar, with $1/r^3$ spatial dependence ($r$ is the interatomic distance), or Van-der-Waals-like, with $1/r^6$ spatial dependence. The collective quantum stochastic resonance studied in the main text persists even when considering the spatial dependence of the Rydberg interactions. To demonstrate this point, we study a one-dimensional chain of Rydberg atoms with $1/r^6$ interaction, where the many-body effects are much weaker than the dipolar type.

\begin{figure*}[h]
\includegraphics[width=0.8\textwidth]{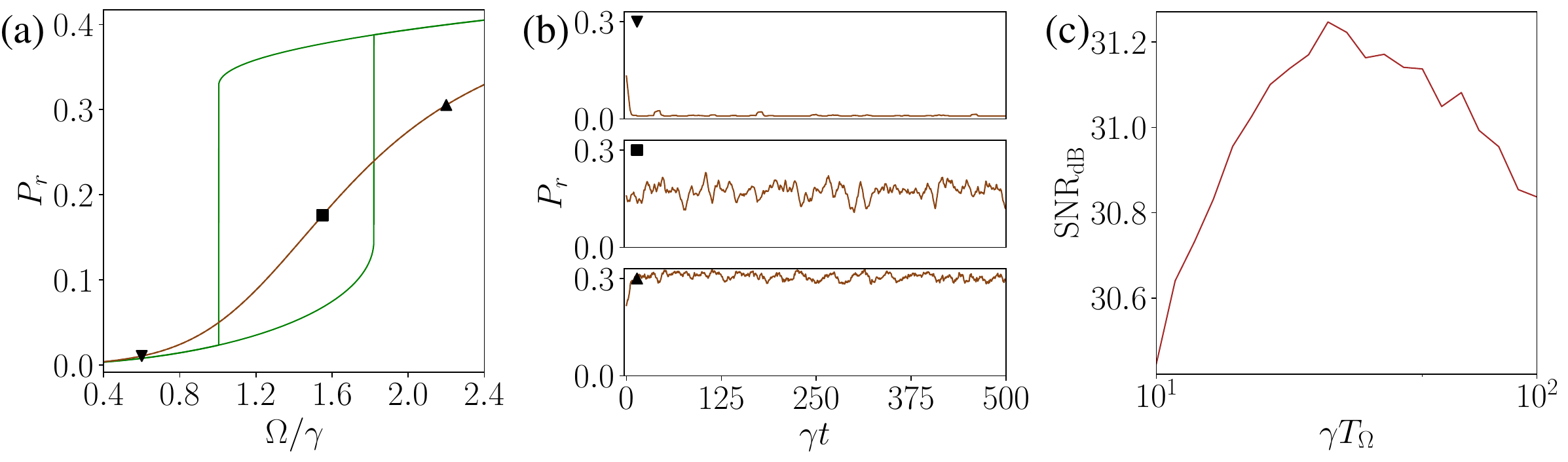}
		\caption{ Stochastic resonance with spatially dependent Rydberg interactions. (a) Rydberg-state population $P_r$ in the steady state of the master equation (brown), and in the bistable states of the optical Bloch equations (green). (b) Single-trajectory dynamics. From top to bottom: $\Omega/\gamma=0.6,1.55,2.2$, as indicated by the corresponding markers in (a). (c) SNR as a function of the modulation period with $\Omega_0/\gamma=1.55$ and $A_\Omega/\gamma=0.2$, averaged over 100 trajectories each with an evolution time $\gamma t=10000$. For all figures, we take $N=8$, with other parameters the same as those in Fig.~1 of the main text.
		}
		\label{figS1}
\end{figure*}

The interaction Hamiltonian is written as
\begin{align}
		H_{int}=\frac{945(N-1)V}{8\pi^6(j-k)^6} \sum_{j<k}(\sigma_j^z+1)  (\sigma_k^z+1),
		\label{Hintqr6}
\end{align}
where we take a dimensionless lattice constant. To facilitate comparison with the mean-field results, the coefficients are determined such that the non-linear detuning term in the optical Bloch equation [Eq.~(4) in the main text] can be derived from Eq.~(\ref{Hintqr6}) once a mean-field approximation is applied.

In Fig.~\ref{figS1}, we show the numerical results for the steady states, the quantum trajectories, and the SNR. The emergence of collective jumps in the middle panel of Fig.~\ref{figS1}(b) and the peaked SNR in Fig.~\ref{figS1}(c) indicate the persistence of the collective quantum stochastic resonance. Compared to the middle panel in Fig.~1(c), the signals for the collective jumps are less prominent here, which we attribute to the reduction in the long-range correlations under the interaction in Eq.~(\ref{Hintqr6}).

\section{Bistability under the cluster model}

The cluster model we introduce in the main text acquires non-linear inter-cluster interactions for $M<N$. This gives rise to bistability-like behaviors at long times, reminiscent of the results under the full mean-field treatment [green curves in Fig.~1(b) of the main text].

\begin{figure*}[tbp]
\centering
\includegraphics[width=0.4\textwidth]{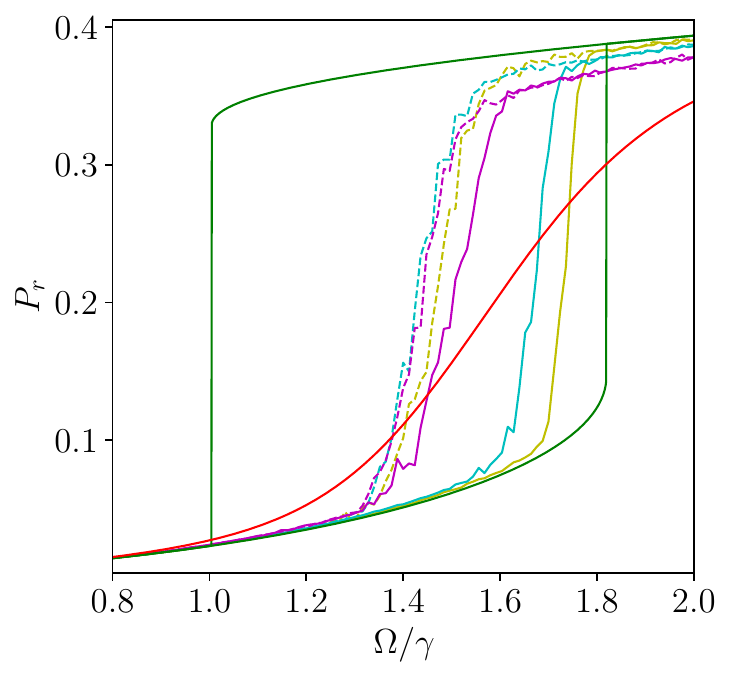}	
		\caption{Bistability with decreasing cluster size $M$.
The population of the Rydberg state $P_r$ is calculated by averaging 100 trajectories, each within the time $\gamma t\in [90,100]$, where the system has relaxed to the vicinity of the steady state. We take $N=8$, and apply the cluster model with $M=4$ (purple), $M=2$ (blue), and $M=1$ (yellow), for two distinct initial Rydberg-state populations: $P_r=0.1$ (solid), and $P_r=0.6$ (dashed).
The red and green curves are the same as those in Fig.~1 of the main text, indicating the steady-state solutions of the full quantum and mean-field models, respectively.
		}
		\label{figS2}
\end{figure*}

To illustrate this point, we numerically evolve the system to the vicinity of its long-time steady state, using the quantum-trajectory scheme under the cluster model.
The ensemble averaged $P_r$ for different initial states are shown in Fig.~\ref{figS2}, which suggests the presence of a bistable region. Specifically, similar to the full mean-field case, initial states with distinct Rydberg populations evolve to final states with distinct Rydberg populations. The hysteresis in the Rydberg population $P_r$ outlines the region with bistability-like behaviors at long times. Clearly, with decreasing cluster size $M$, the bistable region is significantly enlarged, but still lies within the mean-field bistable regime. Combined with the results in Fig.~4 of the main text, we conclude that, with decreasing $M$, the increasingly dominant bistable physics is accompanied by the reduction in the collective-jump rate, both closely related to the gradual breakdown of the long-range quantum correlations.

\section{Classical stochastic resonance under the mean-field model}
In this section, we show that, under the full mean-field model of Eqs.~(3) and (4) in the main text, the collective quantum stochastic resonance does not occur due to the complete lack of collective jumps. Nevertheless, a classical stochastic resonance can be appear if some form of classical noise is introduced. {The underlying mechanism of such a stochastic resonance is fundamentally different from that of the collective quantum stochastic resonance discussed in the main text.}

\begin{figure*}[tbp]
	\centering
	\includegraphics[scale=1]{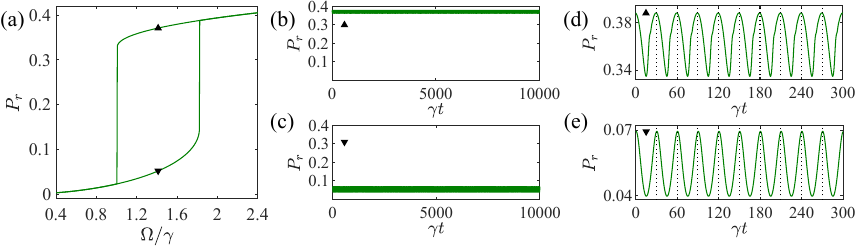}	
	\caption{Dynamics of the Rydberg-state population under the optical Bloch equations, with a periodic drive but without classical noise. (a) Rydberg-state populations in the bistable states. The bistable solutions exist in the range $\Omega/\gamma \in [1.0042, 1.8195]$.
(b)(c) Dynamics of the Rydberg-state population, when the system is initialized in different steady states (indicated by the black markers).
(d)(e) Enlarged view of the dynamics in (b)(c), respectively, at short times.
The parameters are $\Omega_0/\gamma=1.413$, $A_{\Omega}/\gamma=0.39$, and $\gamma T_{\Omega}=30$. Other parameters are the same as those in Fig.~1 of the main text.
	}
	\label{figS3}
\end{figure*}

In Fig.~\ref{figS3}, we show the Rydberg-state population dynamics given by Eqs.~(3) and (4) in the main text, under a periodic modulation of the Rabi frequency $\Omega(t)= \Omega_0 + A_{\Omega}\cos{(2\pi t/T_{\Omega})}$. We choose a $\Omega_0/\gamma=1.413$ within the non-linearity-induced bistable region, as shown in Fig.~\ref{figS3}(a).
In  Fig.~\ref{figS3}(b)(c), we show the time-evolved $P_r=\bar{\rho}_{rr}$ when the density matrix is initialized in the steady state with high (b) an low (c) Rydberg populations, respectively.
Under the periodic drive, the population features oscillations (at the driving frequency) around the initial steady-state value [see Fig.~\ref{figS3}(d)(e)], but there are no signs of collective jumps between the two values. Note that, similar to the condition in the main text (and indeed with the common practice in the classical stochastic resonance), we focus on the weak-signal case, where the periodically modulated $\Omega(t)$ does not cross the boundary of the bistable region. This means that $(\Omega_0\pm A_{\Omega})/\gamma \in [1.0042, 1.8195]$.

We now introduce a noise term to the drive, with
$\Omega(t)= \Omega_0 + A_{\Omega}\cos{(2\pi t/T_{\Omega})}+\sqrt{D}\zeta(t)$.
Here the term $\sqrt{D}\zeta(t)$ describes white noise with zero mean and a variance of $D$. Numerically the white noise is generated by creating a signal consisting of $N$ data points, where each point is sampled from a normal distribution with zero mean of zero and a standard deviation of $\sqrt{D}$. The resulting discrete white-noise signal is then interpolated over the evolution time interval $[0, T_{t}]$ to obtain a continuous function. The range of the noise in the frequency domain is $[0,N/T_{t}]$.
In Fig.~\ref{figS4}(a), we show the time-evolved $P_r=\bar{\rho}_{rr}$ {without any averaging over the white noise.}
The collective state-switching is now visible, as well as the subharmonics in the distribution of inter-jump intervals $p_j$, as we show in In Fig.~\ref{figS4}(b).
{Crucially, the occurrence of the state switching is due to the fact that, with the addition of noise, the system acquires a finite probability to cross both boundaries of the bistable region.
The underlying mechanism is therefore fundamentally different from the emergent collective jumps in the main text.
Again, these emergent collective jumps derive from the interplay of dissipation and Rydberg interactions: they
hinge upon the consecutive quantum jumps of individual atoms, and are further facilitated by the interaction-induced quantum entanglement between atoms.
}

Correspondingly, in Fig.~\ref{figS4}(c)(d), we show the numerically calculated SNR as functions of the driving period and the noise strength $\sqrt{D}$, respectively. In both cases, a peak in the SNR can be identified, signaling the stochastic resonance point. Particularly, the result in Fig.~\ref{figS4}(d) is an exemplary case of the classical stochastic resonance, where the periodic drive and noise conspire to drive the system over the threshold (bistable boundary here), leading to optimized state switching at the resonance point.

\begin{figure*}[tbp]
	\centering
	\includegraphics[scale=1]{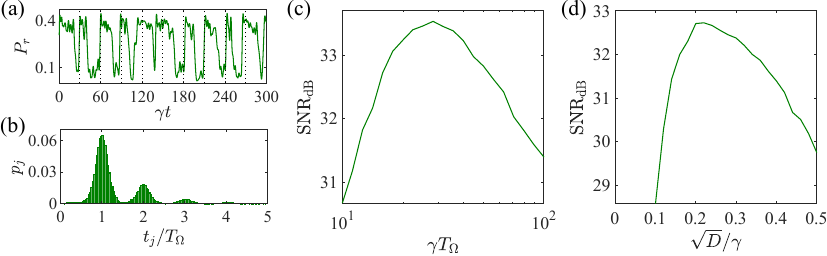}	
	\caption{Classical-noise-induced state switching and stochastic resonance. (a) Dynamics of Rydberg-state population under a periodic drive with white noise. The standard deviation of the noise is $\sqrt{D}/\gamma=0.4$ and the range of the noise in the frequency domain is $[0 ,\gamma]$. (b) Distribution of the inter-jump intervals within the evolution time $\gamma t \in [0,1.6\times 10^6]$. (c) The SNR as a function of $T_{\Omega}$. (d) The SNR as a function of $\sqrt{D}$, for $\gamma T_{\Omega}=100$. The parameters are $\Omega_0/\gamma=1.413$, $A_{\Omega}/\gamma=0.39$, other parameters are the same as those in Fig.~1 of the main text. The curves in both (c) and (d) represent the average of 120 trajectories, each with an evolution time of $\gamma t = 10000$.
	}
	\label{figS4}
\end{figure*}

\section{Different unraveling schemes}

For a given quantum master equation, different unraveling schemes exist that lead to formally different stochastic Schr\"odinger's equations. 
While the ensemble average of the trajectories generated by a stochastic Schr\"odinger's equation is equivalent to the density-matrix dynamics under the master equation, the stochastic state evolution along a single trajectory is only physical (that is, corresponding to a true evolution of the system) when a particular detection scheme is chosen.

For our unraveling scheme in the main text, the corresponding stochastic Schr\"odinger's equation is given as
 \begin{align}
 		d|\psi(t)\rangle=\left[\left(-i H-\frac{1}{2}\sum_{j} L_{j}^{\dagger} L_{j}+\frac{1}{2} \sum_{j}\left\langle L_{j}^{\dagger} L_{j}\right\rangle_t \right) d t+\sum_{j}\left(\frac{L_{j}}{\sqrt{\left\langle L_{j}^{\dagger} L_{j}\right\rangle_t}}-1\right) d N_{j}\right]|\psi(t)\rangle.\label{eq:st1}
 \end{align}
where $\langle \hat{A}\rangle_t:=\langle\psi(t)| \hat{A}|\psi(t)\rangle$, and the random real variables $dN_j$ satisfy
\begin{align}
  		E\left[d N_{j}\right]=\left\langle L_{j}^{\dagger} L_{j}\right\rangle d t, \quad\text{and}\quad d N_{j} d N_{k}=\delta_{jk} d N_{j}.
\end{align}
A direct connection between Eq.~(\ref{eq:st1}) and our unraveling scheme in the main text can be identified, if we require that: i) at any given time, $dN_j$ has
a probability $\left\langle L_{j}^{\dagger} L_{j}\right\rangle_t dt$ to be $1$, otherwise $dN_j=0$; ii) at any given time, at most only one $dN_j=1$. It follows that, the stochastic trajectory of the state evolution under Eq.~(\ref{eq:st1}) is realized by invoking the direct detection of the spontaneously emitted photons~\cite{milburn92}. Under such a detection scheme, $dN_j=1$ corresponds to the detection (at the corresponding time) of a spontaneously emitted photon from the $j$ th atom. According Ref.~\cite{cross12}, the two metastable states along a single trajectory are characterized by the joint emissions (for the plateau with large Rydberg population) and non-emissions (for the plateau with small Rydberg population) of photons, respectively. Therefore, the two plateaus and the collective jumps between them can be probed by recording and counting the rate of spontaneously emitted photons.

	\begin{figure}[tbp]
		\centering
		\includegraphics[width=0.6\textwidth]{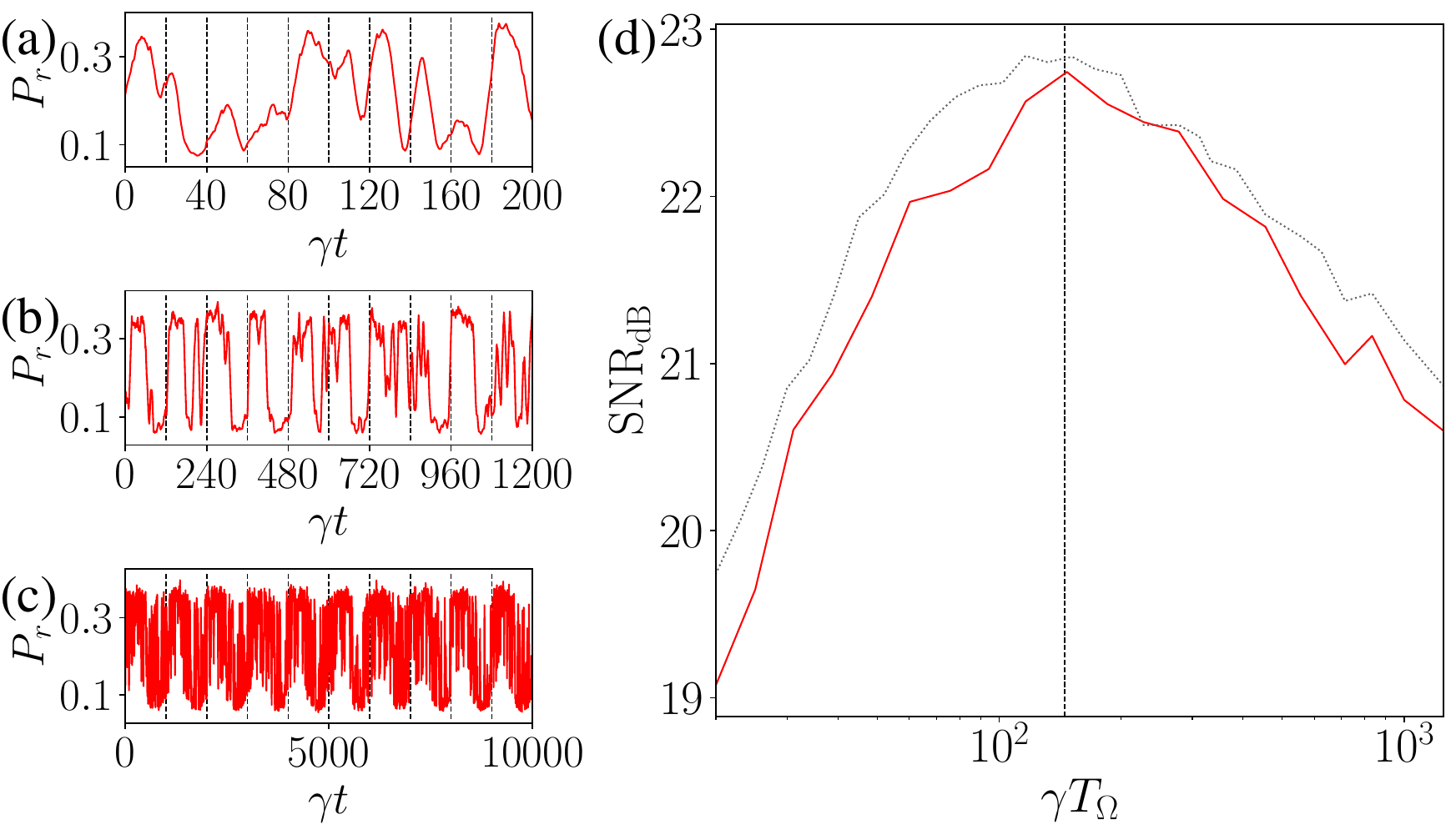}
	\caption{Single-trajectory dynamics and collective quantum stochastic resonance under the Gisin-Percival unraveling. (a)(b)(c) Single-trajectory dynamics of the Rydberg excitation $P_r$ under a periodic drive with $\gamma T_\Omega=20,120,1000$, respectively. (d) Trajectory-averaged SNR under the Gisin-Percival unraveling (solid red). We average over 30 trajectories, and other parameters used for the calculations are the same as those in Fig.~3(a). For comparison, we
reproduce the SNR in Fig.~3(a) of the main text (grey dotted), which corresponds a different unraveling scheme (or detection scheme).
}
		\label{figS5}
	\end{figure}

We now consider a different unraveling scheme, following Gisin and Percival~\cite{gp92p,gp92j}. Unlike the scheme above, it corresponds to diffusive processes of the state vectors.
The corresponding stochastic Schr\"odinger's equation is given as~\cite{milburn92}
\begin{align}
		d\left|\tilde{\psi}(t)\right\rangle=\left\{d t\left[\sum_j\left(\left\langle L_j^{\dagger}\right\rangle_t L_i-\frac{L_j^\dagger L_j}{2}\right)-i H\right]+\sum_jd W_j(t) L_j\right\}\left|\tilde{\psi}(t)\right\rangle,
\end{align}
where $d W_j$ is the increment of a complex Wiener process, satisfying
\begin{align}
		E[d W_j(t)] =0,\quad\text{and}\quad E[d W_j(t)^* d W_j(t)]=dt.
\end{align}
The complex random variable $dW_j$ is generated by two independent, real Gaussian random numbers, with $dW_j=(\xi_1+\mathrm{i}\xi_2)/\sqrt{2}$ and
$\xi_1,\xi_2\sim N(\mu=0,\sigma^2=dt)$. The stochastic trajectory of such an unraveling can be realized by invoking the heterodyne detection of the spontaneously emitted photons~\cite{milburn92}.
In Fig.~\ref{figS5}, we show the single-trajectory dynamics and the trajectory-averaged SNR under the Gisin-Percival unraveling. Apparently, both the collective-jump dynamics and the peaking of the SNR persist. In particular, the SNR peaks at roughly the same location for the two unraveling schemes,
which strongly suggests that both the collective jumps and the phenomenon of the collective quantum stochastic resonance are independent of unraveling schemes. Physically, these emergent phenomena
are universal and can be probed through the counting statistics under different detection schemes. We leave more systematic investigations of the apparent universality of the counting statistics to future studies.

\section{Finite-size effect}

Since our numerics are limited to finite-size systems, it is important to understand the impact of finite-size effect on the stochastic resonance. In Fig.~\ref{figS6}, we show the numerically calculated
trajectory-averaged SNRs for atom numbers $N=8,10,12,14$, respectively. While we have to reduce the number of averaging trajectories that we use for larger $N$ (which leads to more significant fluctuations), it is clear that: i) the peak SNR location remains roughly unchanged; ii) with a larger $N$, the resonance is enhanced, in the sense that the slope on the left side of the peak becomes steeper. Therefore, with larger atom numbers, the resonance peak in the SNR is easier to detect by increasing the driving period from zero.

 	\begin{figure}[tbp]
		\centering	
		\includegraphics[width=0.4\textwidth]{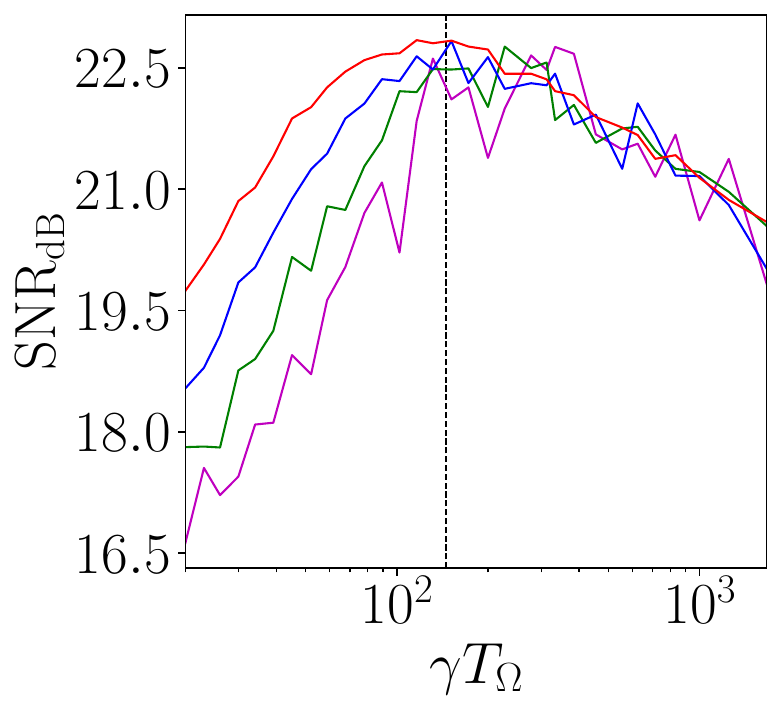}
	\caption{Analysis of the finite-size effect.
The trajectory-averaged SNR for $N=8$ (red),  $N=10$ (blue), $N=12$ (green), and $N=14$ (violet), respectively. The corresponding SNRs are averaged over $100$, $25$, $15$ ,and $4$ trajectories, respectively. Other parameters are the same as those in Fig.~3(a) of the main text.
}
		\label{figS6}
	\end{figure}

\end{widetext}